\documentclass{epl}
\usepackage{psfig}

\title{Lyapunov exponents as a dynamical indicator of a phase transition}
\author{Julien Barr\'e \and Thierry Dauxois}
\institute{Laboratoire de Physique, UMR CNRS 5672, Ecole Normale 
Sup\'erieure de Lyon\\
46, all\'ee d'Italie, 69364 Lyon c\'edex 07, France}
\date{\today}

\begin{document}

\maketitle

\begin{abstract}
We study analytically the behavior of the largest Lyapunov exponent
$\lambda_1$ for a one-dimensional chain of coupled nonlinear
oscillators, by combining the transfer integral method and a
Riemannian geometry approach. We apply the results to a simple model,
proposed for the DNA denaturation, which emphasizes a first order-like
or second order phase transition depending on the ratio of two length
scales: this is an excellent model to characterize $\lambda_1$ as a
dynamical indicator close to a phase transition.
\end{abstract}

\date{\today}

\section{Introduction}
\label{sec:introduction}

{\em Phase transition} plays a central role in equilibrium and non
equilibrium statistical physics books and lectures in particular
because it exemplifies the paradigmatic concept of universality in
physics. In the theory of dynamical systems, the concept of {\em
Lyapunov exponent} has also attracted a lot of
attention~\cite{manneville,lichtenberg} because it defines
unambiguously a sufficient condition for chaotic instability, but
unfortunately except for very few systems it is already an extremely
difficult task to derive analytically the expression of the largest
one, $\lambda_1$, as a function of the energy density.  As some
promising results have been recently obtained to describe some
properties of high-dimensional dynamical
systems~\cite{livi,Constantoudis,ruffo,firpo}, by combining tools
developed in the framework of dynamical systems with concepts and
methods of equilibrium statistical mechanics, the idea that both
concepts could be related was proposed recently~\cite{pettini}.

During the last years, the process by which two strands of DNA
unbind upon heating, called DNA denaturation or melting, has
motivated a lot of works~\cite{simona,grassberger,mukamel,monthus}
and in particular an extremely simplified dynamical
model~\cite{theodorakopoulos} was proposed and studied. This
one-dimensional hamiltonian has the following expression
\begin{equation}\label{Hamiltonian}
 \mathbf{H}=\sum_n \frac{m}{2}{\dot y}_n^2+\frac{K}{2}
 \left(y_n-y_{n+1}\right)^2+D\left(e^{-ay_n}-1\right)^2
\end{equation}
where  $m$ corresponds to the effective mass of nucleotides,  $K $
the coupling constant and $ D$ (resp. $ a $) the depth (resp.
inverse length scale) of the Morse potential which mimic the
interactions between groups of atoms of opposite strands.

This potential with only nearest-neighbor interactions is well suited
for using the transfer integral method in order to derive the
canonical partition function. It can be shown~\cite{theodorakopoulos}
that this system can be mapped to the quantum mechanical analogy of a
particle in a Morse potential and that all thermodynamical quantities
could finally be expressed as functions of the eigenvalues
$\varepsilon_q$ and eigenfunctions $\phi_q$ of this Schr\"odinger
problem. In particular, we showed the existence of a critical
temperature $T_c=2\sqrt{2KD}/ak_B$ corresponding to a second order
phase transition between the so-called native (or double strand) state
with the particles in the bottom of the Morse well, and the
denaturated state, with the particles on the Morse plateau.

The dynamical properties of this one-dimensional model are also
very interesting and have, in particular, emphasized the role of
localized oscillating excitations, called discrete breathers, as
precursors effects driving this phase transition. Spatiotemporal
studies of the dynamics reveals also intermittency-like features
and has led us to consider the importance of its chaotic
properties as an important ingredient to characterize and to
explain  this dynamical instability. This is the reason why it is
important to study the Lyapunov behavior as a function of the
energy density, not only by computing it numerically but also (if
possible !) by deriving its expression using the Riemannian
geometry approach proposed recently~\cite{pettini}.

\section{Riemannian Geometry Approach}
\label{sec:riemman}

The main idea is that the  chaotic hypothesis is
at the origin of the validity of equilibrium statistical physics,
and this should be traced somehow in the dynamics and therefore in
the largest Lyapunov exponent. Through a reformulation of
Hamiltonian dynamics in the language of Riemannian
geometry~\cite{pettini}, the method proposes therefore to relate
the microscopic dynamics to the statistical
averages~\cite{manifold}. Applied to a typical high-dimensional
hamiltonian system, this corresponds to study the parametric-like
instability of geodesics, driven by the fluctuations of the
curvature. Indeed, chaos can be induced  not only by negative
curvatures but also by positive ones, provided they are
fluctuating, as it is less well known. Once the curvature
$\kappa_0 $ and its fluctuations $\sigma_\kappa$ are accurately
determined, the method uses a gaussian stochastic process to
approximate the effective curvature felt along a geodesic and
finally one ends up with the following expression of the largest
Lyapunov exponent
\begin{equation}\label{exprlyapunov}
\lambda_1=\frac{1}{2}\left(\Lambda-\frac{4\kappa_0}{3\Lambda}\right)
\quad\mbox{where}\quad\label{biglambda}
\Lambda=\left(\sigma_\kappa^2\tau+\sqrt{\left(\frac{4\kappa_0}{3}\right)^3
+\sigma_\kappa^4\tau^2}\right)^{1/3}\quad
.
\end{equation}
In this definition, $\tau$, the relevant time scale associated to
the stochastic process is  function of the two following
timescales: $\tau_1\simeq{\pi}/{2\sqrt{\kappa_0+\sigma_\kappa}}$ is the
time needed to cover the distance between two successive conjugate
points along the geodesics, whereas
$\tau_2\simeq{\sqrt{\kappa_0}}/{\sigma_\kappa}$ is related to the local
curvature fluctuations. The general rough physical estimate
${\tau}\simeq\left({1}/{\tau_1}+{1}/{\tau_2}\right)^{-1}$
completes finally the analytical estimate of $\lambda_1$ that we
would like to continue now by the calculation of the mean value of
the curvature and its fluctuations as a function of the energy
density.

\section{Curvature and fluctuations along geodesics}
\label{sec:curvature}

\begin{figure}
\hskip3.5truecm\includegraphics[width=14truecm,height=7truecm]{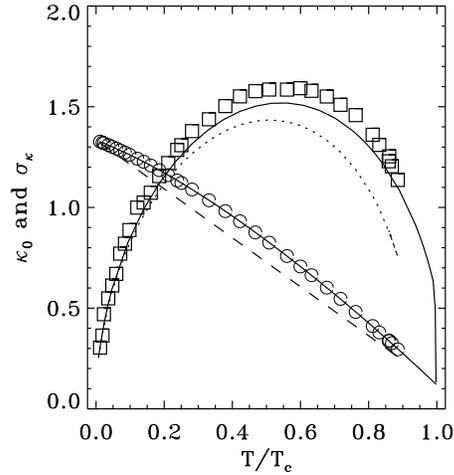}
 \caption{Evolution of $\kappa_0$ (circles) and $\sigma_\kappa$ (squares) as a
function of the temperature $T$. Circles and squares correspond to
microcanonical results, the solid lines to results obtained with
the transfer integral method; the dashed line to the analytical
expression~(\ref{curvature}). Finally, the dotted line correspond
to the low temperature approximation~(\ref{sigmapresde0}) of the
fluctuations of curvature. } \label{gyandgydeux}.
\end{figure}

Using the simplifying Eisenhart metric, the Ricci curvature
$\mathbf{K}_{R}(\mathbf{y})$ corresponds to the Laplacian
of the potential energy. Defining the function
$g(y)=2e^{-2ay}-e^{-ay}$, we get the following expression
\begin{eqnarray}
\kappa_0 ={\langle \mathbf{K}_R\rangle_{\mu} \over N-1} \simeq2K+2
a^2 D \langle g(y)\rangle_\mu
&=&2K+2a^2D\left(1-\frac{T}{T_c}\right) \hskip 1.5truecm \mbox{if}
\quad T<T_c, \label{curvature}\\
&=&2K \hskip 4.5truecm \mbox{if}\quad T >T_c
\end{eqnarray}
where we have used the equality of mean values in microcanonical
$\langle\bullet\rangle_\mu$ and canonical
$\langle\bullet\rangle_{can}$ ensembles~\cite{lebowitz}. The last
expression is in particular derived using the expression of the
groundstate $\phi_0(y)$ of the quantum mechanical analogy since
$\langle
g(y)\rangle_{can}=\langle\phi_0|g(y)|\phi_0\rangle_{can}$.
Figure~\ref{gyandgydeux} attests that the agreement is excellent
between the above analytical expression, the canonical transfer
integral results and microcanonical molecular dynamics
simulations, obtained using the best 4th order symplectic
integrator due to MacLahan-Attela~\cite{parameters}.

It is interesting to note that, the mean curvature is always
positive  whereas the local one is negative close to the inflexion
point of the Morse potential. In addition,
expression~(\ref{curvature}) emphasizes the role of discreteness
since the curvature would be almost constant in the continuum
limit $a^2D\ll K$. This will have further consequences on the
evolution of $\lambda_1$ when the discreteness is changed.

\section{Fluctuations of curvature}
\label{sec:fluctuations}

Since numerical computations are simpler in the microcanonical
ensemble, while analytical calculations are simpler in the
canonical one, to get the fluctuations of the curvature in the
microcanonical ensemble, one needs to add the following corrective
term~\cite{lebowitz}
\begin{eqnarray}
\langle \delta^2 K_R\rangle_{\mu} &=& \langle \delta^2
K_R\rangle_{can} -\left(\frac{\partial U}{\partial
\beta}\right)^{-1} \left(\frac{\partial \langle K_R
\rangle_{can}}{\partial\beta}\right)^2\quad.
\end{eqnarray}
The expression of the energy
\begin{eqnarray}
\frac{U}{N}=   \frac{1}{2\beta}-\frac{1}{N}
\frac{\partial\ln Z_{can}}{\partial \beta}
&=&k_BT+ D \ \frac{T^2}{T_c^2} \hskip 1.5truecm \mbox{if}\quad
T<T_c, \label{uverT}\\
 &=&k_BT \hskip 2.8truecm \mbox{if}\quad
T>T_c, \label{uverTsurtc}
\end{eqnarray}
emphasizes the second order of the phase transition, and finally
using the expression of the mean value of the curvature, we obtain
the microcanonical/canonical correction
\begin{eqnarray}
\langle\delta^2K_R\rangle_{\mu}-\langle\delta^2K_R\rangle_{can} &
=&-\frac{4a^4D^2N}{1+\sqrt{\frac{a^2D}{ 2K}}{T\over T_c}}\
\frac{T^2}{T_c^2} \label{correcmicro}
\end{eqnarray}
valid below $T_c$. This quadratic correction vanishes for very low
temperature and converges toward a negative constant close to
$T_c$.

The expression of the fluctuations of curvature in the canonical
ensemble is then
\begin{eqnarray}
\langle\delta^2K_R\rangle_{can}=\langle
\mathbf{K}_{R}^2(\mathbf{y})-\langle
\mathbf{\mathbf{K}}_{R}(\mathbf{y})\rangle^2\rangle_{can}
 & = & 4a^4D^2\sum_{i,j}\biggl(\langle g(y_{i})g(y_{j})\rangle_{can}-
\langle g(y_{i})\rangle_{can} \langle
 g(y_{j})\rangle_{can}\biggr).
\nonumber\end{eqnarray} The second part of the parenthesis could
be easily computed and gives a term $N^2\langle g\rangle^2$. The
first one can be evaluated with the transfer integral method
since, thanks to the periodic boundary condition, one needs to
take into account only the difference in indices $(i-j)$. However,
one cannot integrate directly the integral over the resulting
$(N-1)$ variables, and one needs to decompose twice on the
orthonormal basis $\{\phi_q\}$ of the transfer operator. Defining
the matrix elements of $g(y)$ in this base
\begin{equation}
  M_{k,q_0}=\int_{-\infty}^{+\infty}\phi_{q_0}^\ast(y)g(y)\phi_{k} (y)dy=
\langle\phi_{q_0}|g(y)|\phi_{k}
  \rangle\quad,
\end{equation}
we finally obtain
\begin{eqnarray*}
\langle g(y_{p})g(y_{N})\rangle_{can}& = & \frac{1}
{Z_{c}}\sum_{k,q}  e^{-
(N-p)\beta(\varepsilon_{q}-\varepsilon_{k}) } \ e^{-
N\beta\varepsilon_{k}}\ \ |M_{k,q}|^2\quad.
\end{eqnarray*}
As  only the lowest eigenvalue $\varepsilon_{0}$ will contribute
in the thermodynamic limit, we finally get
\begin{eqnarray}
\langle\delta^2K_R\rangle_{can} & = &  4a^4D^2N
\sum_{q=1}^{+\infty}
\frac{\left|M_{0,q}\right|^2}{1-e^{-\beta(\varepsilon_{q}-\varepsilon_{0})
}} \label{fluccourbure}
\end{eqnarray}
by recognizing a geometrical sum. Figure~(\ref{gyandgydeux})
emphasizes the excellent agreement between microcanonical
simulations (squares) and the above analytical expression (solid
line) where we have used the eigenfunctions of the transfer
integral operator.

\section{Low temperature approximation}
\label{lowT}
 The above expression~(\ref{fluccourbure}) is however
difficult to handle in general but could be simplified in the low
temperature regime since the gap between the eigenvalue of the
ground state $\varepsilon_0$ and the other ones justify to neglect
the exponential term in the denominator. This approximation breaks
down of course close to the phase transition since the critical
temperature is defined by the disappearance of the ground state
$\varepsilon_0$ in the continuum. We obtain thus
\begin{eqnarray}
\langle\delta^2K_R\rangle_{can} &\simeq&4a^4D^2N
\sum_{q=1}^{+\infty}   \left|\langle
\phi_{0}|g|\phi_{q}\rangle\right|^2=4a^4D^2N \left[   \langle
g^2\rangle-\langle g\rangle^2 \right]\quad.
\end{eqnarray}
Using the procedure described for the calculation of $\langle g
\rangle$, one obtains the formula $\langle g^2
\rangle=\left(1-\frac{T}{T_c}\right)\left(1+\frac{4T}{T_c}
+\frac{2T^2}{T_c^2}\right) $, where the prefactors 2 and 4 have been
renormalized to be in agreement with numerical simulations; the
discreteness of the chain, which drives the phase transition,  
is the main reason for this renormalization. Adding finally
the microcanonical correction~(\ref{correcmicro}), we obtain
\begin{eqnarray}
  \sigma_\kappa^2&\equiv&\frac{\langle\delta^2K_R\rangle_{\mu}}{N-1}=4a^4D^2 
\left[
\left(1-\frac{T}{T_c}\right)\left(\frac{5T}{T_c}+\frac{2T^2}{T_c^2}\right)
-\frac{1}{1+\sqrt{\frac{a^2D}{ 2K}}{T\over T_c}}\ \frac{T^2}{T_c^2}\right]
\quad.
\label{sigmapresde0}
\end{eqnarray}
 The dotted
line plotted in Fig.~(\ref{gyandgydeux}) attests the quality of
this low temperature approximation which leads to a linear
dependence versus temperature of the fluctuations $
\sigma_\kappa^2\sim 20a^4D^2 {T}/{T_c}$ at the lowest order.

In the neighborhood of the phase transition, the last bound state
is too close to the continuum: the exponential term in the
denominator of Eq.~(\ref{fluccourbure}) cannot be  neglected and
one needs to take into account fully the interactions with the
continuum. Due to the rapid oscillations of the scattering
eigenstates, one could use the approximate expression for the
phase shift of the scattering states, as it was successfully shown
in the calculation of the static structure factor and therefore
for the susceptibility as its limit~\cite{theodorakopoulos}. But,
unfortunately for the present calculation, there is  an important
contribution of eigenstates with nonvanishing eigenvectors which
prevents this high temperature approximation to succeed.

\section{Largest Lyapunov exponent}

\begin{figure}
\null\hskip 1truecm\twofigures[width=6truecm,height=6truecm]{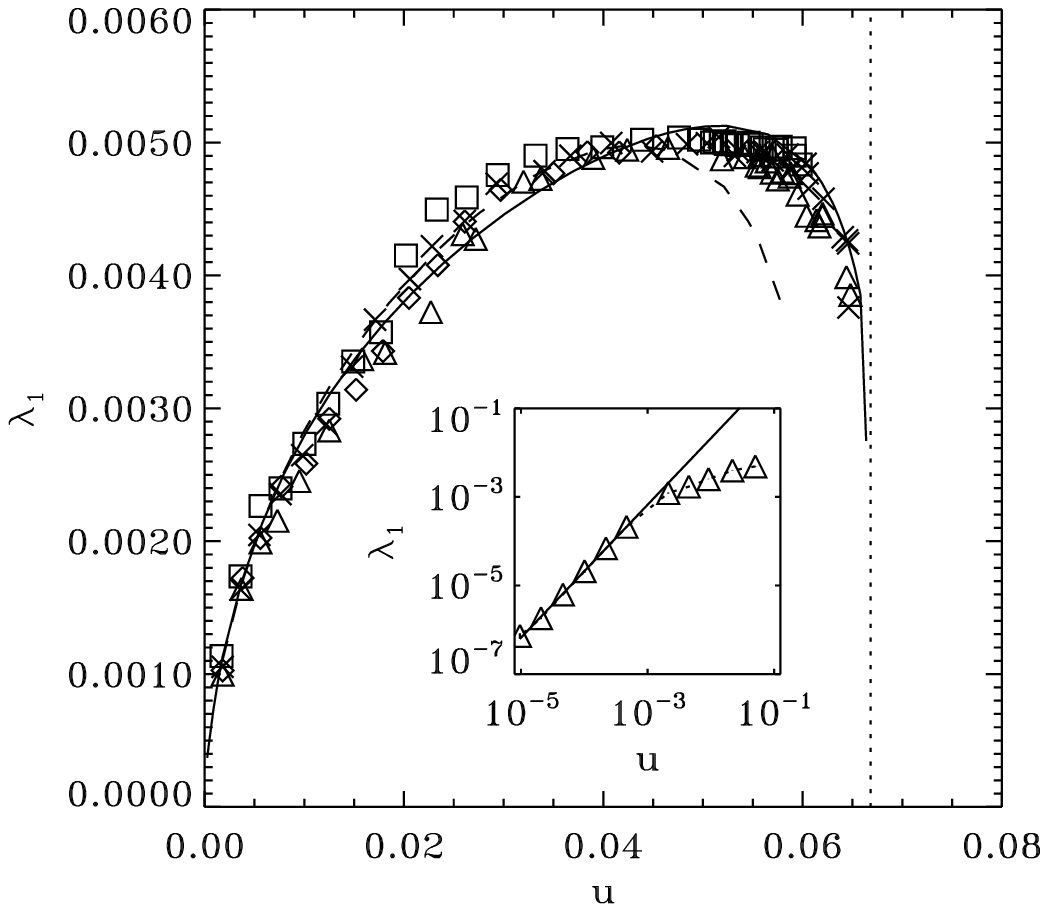}{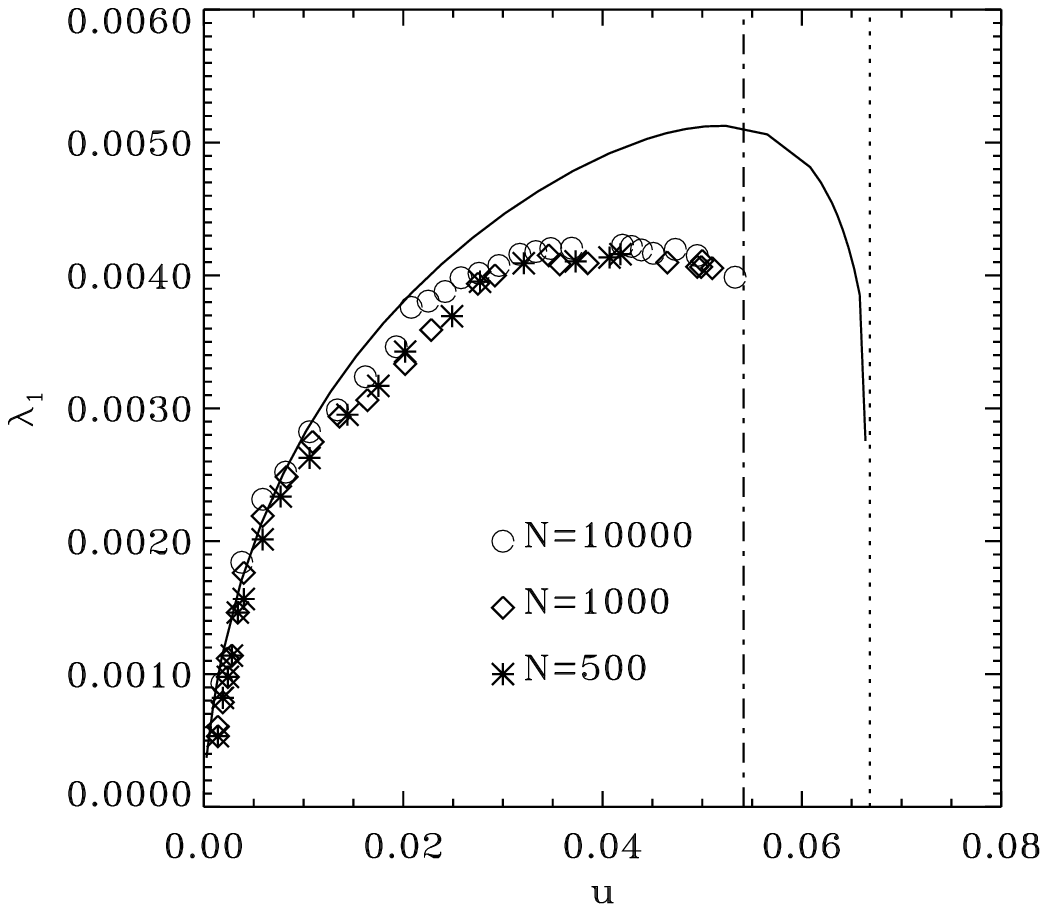}
\caption{Evolution of the maximal Lyapunov exponent as a function
of the energy density $u=U/N$. The symbols correspond to
microcanonical numerical simulations for $N=$500 (triangles),
$N=10^3$ (squares), $N=2.10^3$ (crosses) and $N=10^4$
(diamonds). The solid line corresponds to the analytical
estimation using the transfer integral method whereas the dashed
line corresponds to the analytical expression valid at low energy.
The vertical dotted line shows the position of the critical energy
density $u_c=k_BT_c+D$. In the inset, we plot $\lambda_1$ as a
function of the energy density $u$ in log-log scale. The symbols
correspond to microcanonical numerical simulations for $N=500$
whereas the solid line to a 3/2 power law.} \label{lyapfinal}
\caption{Lyapunov exponent  and order of the phase transition. 
The solid line shows the evolution of the
harmonic model with second order phase transition, whereas the symbols
are referring to the model with anharmonic coupling with a first-order
like phase transition.  The number of sites in the chain is $N=$500
(crosses), $N=10^3$ (diamonds) and $N=10^4$ (circles). The vertical
dotted (resp. dash-dotted) line corresponds to the critical energy
density of Hamiltonian with harmonic (resp. anharmonic) coupling.}
\label{doublelyap}
\end{figure}

Once the curvature~(\ref{curvature}) and its
fluctuations~(\ref{fluccourbure}) have been calculated, it is
straightforward to compute the two timescale $\tau_1$ and $\tau_2$.
Having in mind that the estimate of the decorrelation time scale
$\tau$ is still somehow rudimentary~\cite{pettini} in the Riemannian
framework outlined above, Fig.~(\ref{lyapfinal}) shows that the
alternative definition
${\tau}\simeq\left({1}/{\tau_1}+{5}/{\tau_2}\right)^{-1}$ gives a
particularly striking agreement between the microcanonical numerical
results obtained using the standard algorithm~\cite{benettin} and the
analytical derivation proposed here. Let us emphasize that the
agreement is excellent on the whole interval contrary to earlier
results obtained for the $\phi^4$ chain~\cite{caiani}. The approach
described here has a domain of validity even larger than expected
since the agreement is very good even in the region where the
fluctuations of curvature are of the order or greater than the
curvature itself.

It is also very interesting to remark that the simultaneous use of the
Riemannian geometry approach and the transfer integral method, that we
propose here, could be easily extended to the treatment of any chain
of oscillators with on-site potential and nearest-neighbor harmonic
coupling. Once the transfer integral operator is solved analytically
as it is possible here for Hamiltonian~(\ref{Hamiltonian}) or
numerically in general, the evolution of the largest maximal Lyapunov
exponent as a function of the energy density is directly derived from
expressions of the curvature~(\ref{curvature}) and
fluctuations~(\ref{fluccourbure}), by simply replacing $2a^2Dg(y)$ by the 
second derivative of the onsite potential.

A closer look at very low energy density, shows however a non
surprising disagreement since the inset plotted in
Fig.~(\ref{lyapfinal}) clearly violates the low energy density
approximate law $\lambda_1\sim\sigma_\kappa^2$, which would give
here a linearly increasing function of the temperature according
to the lowest order estimate of $\sigma_\kappa$. The reason is
presumably the breakdown of the stochastic approximation for the
average curvature in this very low energy limit where the system
is almost quadratic and therefore the crucial ergodic hypothesis
is not completely fullfilled any more.  More interestingly, the
numerical model emphasizes an unusual 3/2 power law contrary to
the generally reported $u^2$ dependence of $\lambda_1$ in
high-dimensional dynamical systems~\cite{pettini,firpo}.

Let us note that the expression of $\kappa_0$ is strongly dependent of
the discreteness parameter $a^2D/K$, but the fluctuations
$\sigma_\kappa$ are not. As Fig.~(\ref{lyapfinal}) shows that, away
from the critical region, smaller the curvature is, greater the
Lyapunov is, one can expect that the more discrete the chain is, the
more chaotic its dynamics would be. This result is clearly in
qualitative agreement with the discovery that spontaneously created
discrete breathers are actually {\em chaotic} excitations and that
their domain of stability is far from the continuum
limit~\cite{aubry}.

\section{Lyapunov and Phase transition}
This model is also a very interesting case where one can characterize
$\lambda_1$ as a dynamical indicator of a  phase
transition.  In this case of a second order phase transition, one
obtains a critical slowing down reminiscent of the results obtained
for the Heisenberg Mean-field model derived by Firpo~\cite{firpo}. We
would like also to emphasize the singular behavior of $\kappa_0$,
$\sigma_\kappa$ and $\lambda_1$ at the critical energy density, in
complete agreement with the exciting conjectures~\cite{pettini}
linking them to a topology change in the underlying manifold, being
itself an indicator of a thermodynamic phase transition.

 More interestingly, we also have studied
 Hamiltonian~(\ref{Hamiltonian}) whith an anharmonic coupling
 potential ${K\over 2} [1+ e^{-\alpha(y_n+y_{n-1})} ] \
 (y_n-y_{n-1})^{2}$ which has the property to describe the varying
 backbone stiffness of the DNA. This hamiltonian was shown to have a
 first order like phase transition~\cite{ziaandnikos} when the ratio
 of the two inverse length scales $\alpha/a$ is lower than $1/2$.  If
 analytical estimate of the Lyapunov exponent are not possible,
 Fig.~\ref{doublelyap} shows its evolution obtained using
 microcanonical molecular dynamics simulations.  In the low energy
 limit, the additional parenthesis do not produce of course
 modifications, but it emphasizes an abrupt change~\cite{MEHRA} of the
 Lyapunov exponent at the critical energy density as if $\lambda_1$
 was a dynamical order parameter indicating the {\em first} order
 phase transition.

\section{Conclusion}

In this letter, we have presented one of the very few analytical
calculation of the largest Lyapunov exponent in a high-dimensional
dynamical system~\cite{lichtenberg}. As expected, in addition to
serve as a criterion for chaos, or as a characteristic time scale
of chaoticity, $\lambda_1$ is an excellent dynamical indicator of
the presence of the phase transition and could be used not only to
describe the dynamics but also the statistical properties of high
dimensional systems.

It is also interesting to consider the behavior of another
important dynamical indicator, the {\em participation ratio} of
the normalized Lyapunov vector $V_1$ associated to the maximal
Lyapunov $\lambda_1$. Defined as  $\xi={1}/
\left(N\sum_{i=1}^N\left[V_1(i)^2+ V_1(i+N)^2\right]^2\right)$,
where the first (resp. last) $N$ components  are associated to the
evolution of linear perturbations of $y_n$ (resp. ${\dot y}_n$) in
tangent space, $\xi$ is an indicator of
localization~\cite{pikovskypoliti,julienchate}: it is of order one
if the vector is extended and of order $1/N$ if localized. Here,
we found that the phase transition corresponds to a crossover from
a localized state in tangent space at low energy density to a more
extended state just after the phase transition,
confirming that the tangent space trough the largest Lyapunov
exponent $\lambda_1$, but also its associated eigenvector $V_1$
through the participation ratio $\xi$, are surprisingly good
dynamical indicators for emphasizing thermodynamical phase
transitions.

\acknowledgments We wish to acknowledge very helpful discussions with
A. Alastuey, J. Farago, M.-C. Firpo, S. Lepri, R. Livi, M.
Peyrard, S. Ruffo, M. Terraneo, N. Theodorakopoulos.


\begin{thebibliography}{0}
\bibitem{manifold}The natural motion can be viewed as geodesics of
a suitable Riemannian manifold.
\bibitem{parameters}
The parameters of the model are
$D=0.03$ eV, $K=0.06$ eV/$\AA^2$, $a=4.5\ \AA^{-1}$,
$\alpha=0.35\ \AA^{-1} $ 
and $m=300$ a.m.u.
\end{thebibliography}
\end{document}